\documentclass[a4paper]{amsart}
\usepackage{latexsym}
\usepackage{amsmath}
\usepackage{amssymb}
\usepackage{amsfonts}
\usepackage{hyperref}
\usepackage{color}
\usepackage{enumitem}
\definecolor{Blue}{rgb}{0.,0.,1.}
\definecolor{Red}{rgb}{1.,0.,0.}

\newcounter{smallarabics}
\newenvironment{arabicenumerate}
{\begin{list}{{\normalfont\textrm{(\arabic{smallarabics})}}}
  {\usecounter{smallarabics}\setlength{\itemindent}{0cm}
   \setlength{\leftmargin}{5ex}\setlength{\labelwidth}{4ex}
   \setlength{\topsep}{0.75\parsep}\setlength{\partopsep}{0ex}
   \setlength{\itemsep}{0ex}}}
{\end{list}}

\newcounter{smallroman}

\newcommand{\ben}{\begin{arabicenumerate}}  
\newcommand{\een}{\end{arabicenumerate}}

\newtheorem{theoreme}{Theorem }[section]
\newtheorem{proposition}[theoreme]{Proposition}
\newtheorem{lemma}[theoreme]{Lemma}
\newtheorem{definition}[theoreme]{Definition}

\newtheorem{remark}[theoreme]{Remark}
\newtheorem{example}[theoreme]{Example}
\def\beq#1\eeq{\begin{align}#1\end{align}}
\def\beqa#1\eeqa{\begin{align}#1\end{align}}
\def\bes#1\ees{\begin{split}#1 \end{split}}

\newcommand{\bex}{\begin{example}}
\newcommand{\eex}{\end{example}}
\def\bel{\begin{lemma}}
\def\eel{\end{lemma}}
\def\bet{\begin{theoreme}}
\def\eet{\end{theoreme}}
\def\bed{\begin{definition}}
\def\eed{\end{definition}}
\def\ber{\begin{remark}}
\def\eer{\end{remark}}

\def\slim{{\rm s-}\lim}

\def\i{{\rm i}}

\def\Sp{{\mathcal Sp}}

\def\bbbone{{\mathchoice {\rm 1\mskip-4mu l} {\rm 1\mskip-4mu l}
{\rm 1\mskip-4.5mu l} {\rm 1\mskip-5mu l}}}
\def\one{\bbbone}

\def\12{\frac{1}{2}}

\def\supp{{\rm supp}}

\def\e{{\mathrm e}}

\def\bep{\begin{proposition}}
\def\eep{\end{proposition}}
\newcommand{\cp}{\mathrm{c}}

\newcommand{\nat}{\mathbb{N}}
\newcommand{\ti}{\tilde}
\newcommand{\w}{\mathrm{w}}

\newcommand{\De}{\Delta}

\newcommand{\mcL}{\mathcal{L}}

\newcommand{\lan}{\langle}
\newcommand{\ran}{\rangle}

\newcommand{\pa}{\partial}

\newcommand{\Om}{\Omega}

\newcommand{\hil}{\mathcal{H}}
\newcommand{\om}{\omega}
\newcommand{\mfa}{\mathfrak{A}}
\newcommand{\mco}{\mathcal{O}}

\newcommand{\fr}[2]{\frac{#1}{#2}}

\newcommand{\real}{\mathbb{R}}

\newcommand{\la}{\lambda}
\newcommand{\ov}{\overline}

\newcommand{\non}{\nonumber}

\newcommand{\wh}{\widehat}

\def\bbbone{{\mathchoice {\rm 1\mskip-4mu l} {\rm 1\mskip-4mu l}
{\rm 1\mskip-4.5mu l} {\rm 1\mskip-5mu l}}}
\def\one{\bbbone}
\def\cS{\mathcal{S}}
\def\supp{{\rm supp}}

\def\12{\frac{1}{2}}

\usepackage{verbatim}

\DeclareFontFamily{U}{mathx}{\hyphenchar\font45}
\DeclareFontShape{U}{mathx}{m}{n}{
      <5> <6> <7> <8> <9> <10>
      <10.95> <12> <14.4> <17.28> <20.74> <24.88>
      mathx10
      }{}
\DeclareSymbolFont{mathx}{U}{mathx}{m}{n}
\DeclareFontSubstitution{U}{mathx}{m}{n}
\DeclareMathAccent{\widecheck}{0}{mathx}{"71}
\DeclareMathAccent{\wideparen}{0}{mathx}{"75}

\newlength{\dinwidth}
\newlength{\dinmargin}
\begin{document}

\title[The problem of asymptotic completeness in AQFT]{Asymptotic observables, propagation estimates and the problem of asymptotic completeness in algebraic QFT}
\author{Wojciech Dybalski}
\address{ Institut f\"ur Theoretische Physik,  ETH Z\"urich, 8093 Z\"urich
Switzerland 
and 
Zentrum Mathematik, Technische Universit\"at M\"unchen,
D-85747 Garching Germany}
\email{dybalski@ma.tum.de}


\begin{abstract} 

We review recent results on the existence of asymptotic observables  in algebraic QFT. The problem of asymptotic completeness is discussed from this perspective.

\end{abstract}
\maketitle

\section{Introduction}

The problem of existence of asymptotic observables has a long
tradition in algebraic quantum field theory (AQFT), starting with a seminal work of Araki and Haag \cite{AH67}. 
These authors introduced certain time-dependent families of observables, and showed that their limits  as time tends to 
infinity
(\emph{i.e.} certain asymptotic observables), behave as idealized  particle detectors\footnote{We consider only the case of outgoing asymptotic observables, since the incoming case is analogous.}.  However, 
in \cite{AH67}  the convergence  was only shown on certain domains of scattering states.  
The existence of such asymptotic observables on arbitrary vectors of bounded energy has remained an open problem for over four
decades. It can be expected from quantum mechanics that a solution of this problem is a key to asymptotic completeness in AQFT.

In an ongoing project with C. G\'erard  we gave a solution of this problem for a certain class of detectors \cite{DG12,DG13}. Moreover, we demonstrated  that the linear span of the ranges of these detectors coincides with the subspace of scattering states.  This weak variant of asymptotic completeness has a simple physical interpretation: given any initial state (think of a container with hydrogen gas used as a proton source at the LHC), after
a typical particle physics experiment one obtains a configuration of independent particles in terms of which the measurement results are interpreted. 
We show that in the context of massive theories this empirical fact can be derived from the basic assumptions of locality and positivity of energy in a model-independent manner.

\section{Framework} \label{Framework}

We work in the standard framework of  AQFT, which is given by:

\begin{enumerate}
\item  A  net of local von Neumann algebras $\mco\mapsto\mfa(\mco)\subset B(\hil)$, labelled by open bounded regions of Minkowski spacetime 
$\mco\subset \real^{d+1}$. $\mfa(\mco)$ is interpreted as the 
algebra of all observables measurable in $\mco$. 

\item  The global algebra of this net $\mfa:=\ov{\bigcup_{\mco\subset \real^{d+1}} \mfa(\mco)}$.
\item A strongly continuous unitary representation of translations $\real^{d+1}\ni (t,x)\mapsto U(t,x)$ acting on $\hil$. Its generators $(H,P)$ have the interpretation of the total energy and momentum operators. Their joint spectrum
is denoted $\Sp\,U$ and their spectral projection on a Borel set $\De$ by $\one_{\De}(U)$.
\end{enumerate}

On these objects we impose the Haag--Kastler postulates:
\begin{enumerate}

\item (isotony) $\mco_1\subset \mco_2  \Rightarrow \mfa(\mco_1)\subset \mfa(\mco_2)$.

\item (locality) $\mco_1\, {\textrm{\Large $\times$} } \, \mco_2\Rightarrow  [\mfa(\mco_1),  \mfa(\mco_2)]=0$, where 
${\textrm{\Large $\times$} }$ means  spacelike separation.

\item (covariance) $U(t,x)\mfa(\mco) U(t,x)^{*}=\mfa(\mco+(t,x))$.

\item (positivity of energy) $\Sp\, U=\{0\}\cup H_m\cup G_{2m}$.

\item (uniqueness and cyclicity of the vacuum) 
$\one_{\{0\}}(U)=|\Om\ran \lan \Om|$ and $\ov{\mfa\Om}=\hil$.

\end{enumerate}

We adopt a restrictive variant of the positivity of energy  assumption (4), suitable for massive theories, which requires that     $\Sp\, U$ consists of $\{0\}$, corresponding to the vacuum vector $\Om$, a mass hyperboloid $H_m:=\{\,(E,p)\in\real^{d+1}\,|\, E=\sqrt{p^2+m^2}\,\}$ carrying 
single-particle states and the multiparticle spectrum $G_{2m}:=\{\,(E,p)\in\real^{d+1}\,|\, E\geq \sqrt{p^2+(2m)^2}\,\}$.  
We recall that there exist interacting quantum field theories which satisfy the above assumptions, for example $\la\phi^4_2$ at small $\la$ \cite{GJS73}.

\section{Generalized creation/annihilation operators}
\setcounter{equation}{0}

In concrete models, the conventional building blocks
of physical observables are creation and annihilation operators.
To construct their counterparts  in our abstract setting,  
we need to control the energy-momentum transfer of
elements of the algebra of observables $\mfa$. The energy-momentum transfer (or the Arveson spectrum) of an observable $B\in \mfa$ is denoted $\supp\,\wh{B}$ and defined as the support of the operator-valued distribution:
\beqa 
\ti B(E,p):= (2\pi)^{ -\fr{d+1}{2} }\int dt dx\, \e^{-\i Et+\i px}B(t,x),
\eeqa
where $B(t,x):=U(t,x)BU(t,x)^*$. It has the
expected properties, in particular
\beqa
\supp\,\wh{B^*}&=-\supp\,\wh{B},\\
B\one_{\De}(U)\hil& \subset \one_{\ov{\De+\supp\,\wh{B}} }(U)\hil. \label{energy-momentum-transfer}
\eeqa
To use $B^*$ as a creation operator of a single-particle state, we would
like $\supp\,\wh{B^*}$ to be a small neighbourhood of a point on the
mass hyperboloid $H_m$.  An operator with such an energy-momentum
transfer cannot be strictly local, \emph{i.e.} contained in $\mfa(\mco)$ for some
open bounded region $\mco$. However, it can be \emph{almost local}
\emph{i.e.} it can be approximated  in norm by operators localized in double-cones, 
centred at zero, with increasing radius, up to an error vanishing faster than any 
inverse power of this radius. A typical example of an almost local operator is
\beqa
B=\int dtdx\, A(t,x)f(t,x), \quad A\in \mfa(\mco), f\in \cS(\real^{d+1})
\label{example-B}
\eeqa
for some open bounded $\mco$. Clearly, $\supp\,\wh{B}\subset\supp\hat f$,
so the energy-momentum transfer of $B^*$ can be a compact set.
Also, for such operators the function $(t,x)\mapsto B(t,x)$ is smooth
in norm and all its derivatives are again almost local.
We denote by $\mcL_0\subset\mfa$ the subspace of operators of the form (\ref{example-B}) 
whose energy-momentum transfers are compact sets supported outside of the future 
lightcone\footnote{This definition is slightly more restrictive than the one from \cite{DG12,DG13}.}. 
The elements of $\mcL_0$ are `annihilation operators' in the sense that they annihilate the vacuum, 
but not in the sense of  canonical commutation relations.

The above construction of generalized creation and annihilation operators
is well known since  early days of AQFT. We conclude this section 
with a more recent concept of `improper' annihilation operators,
introduced in our recent work \cite{DG12}. They are defined as maps 
\beqa
& a_B: \hil_{\cp} \mapsto \hil \otimes L^2(\real^d),\\
& (a_B\Psi)(x)=B(x)\Psi,
\eeqa
where $B\in \mcL_0$ and  $\hil_{\cp}$ is the domain of vectors $\Psi\in \hil$ s.t. $\Psi=\one_{\De}(U)\Psi$ for some compact set $\De$. The fact that $x\mapsto B(x)\Psi$ is square-integrable is non-trivial, but
it follows from Lemma 2.2 of \cite{Bu90}. Similarly, for a family
$B_1,\ldots, B_n$ of operators from $\mcL_0$, we define
\beqa
&a_{B_1,\ldots, B_n}:  \hil_{\cp} \to \hil \otimes L^2(\real^{nd}),\\
& (a_{B_1,\ldots, B_n}\Psi)(x_1,\ldots, x_n)=B_1(x_1)\ldots B_n(x_n)\Psi.
\eeqa  
\section{Propagation observables and asymptotic observables}
\setcounter{equation}{0}

We recall that in non-relativistic scattering theory, time-dependent estimates on the propagation properties of solutions of evolution equations 
are usually called {\em propagation estimates}, and time-dependent observables used to derive them are called {\em propagation observables}. 
Asymptotic observables are limits (as time $t\to\infty$) of propagation observables evolved in the Heisenberg picture.

Let us first consider  the classical phase space $T^*\real^d=\real^d\times (\real^d)'$, $h\in \cS(T^*\real^d)$
and a classical propagation observable 
\beqa
t\mapsto h_t\in \cS(T^*\real^d),
\eeqa
where $h_t(x,\xi):=h(x/t,\xi)$. 
We can elevate it to a quantum-mechanical propagation observable with the
help of the Weyl quantization:
\beqa
 t\mapsto h_t^\w\in B(L^2(\real^d)),
\eeqa
where
\beqa
 (h^{\w}_t u)(x)  = (2\pi)^{-d}\int \e^{\i (x-y)\cdot \xi}h_t\big(\frac{x+y}{2}, \xi\big)u(y)  dy d\xi, \ u\in L^2(\real^{d}).
\label{Weyl-quantization}
\eeqa 
Next, using the map $a_B$, $B\in \mcL_0$, we define a quantum-field-theoretical propagation
observable:
\beqa
t\mapsto  a_{B}^*(\one_{\hil}\otimes h_t^\w)a_B,  \label{propagation-observable}
\eeqa 
which is a family of  (unbounded) operators on $\hil_{\cp}$.

Given the propagation observable (\ref{propagation-observable}), the corresponding asymptotic observable
is approximated (as $t\to\infty$) 
by:
\beqa
C_t:= \e^{\i tH} a_{B}^*(\one_{\hil}\otimes h_t^\w)a_B \e^{-\i tH}. \label{C-t-observable}
\eeqa 
We mention as an aside that in the  case of $h$ independent of momentum $\xi$ the above formula gives
\beqa
C_t=\int dx\, h(x/t) (B^*B)(t,x), \label{Araki-Haag}
\eeqa
which is the standard  Araki--Haag detector \cite{AH67,Bu90}.

Our main result, stated in Theorem~\ref{Main-result} below, concerns the strong convergence  as $t\to\infty$  of 
approximating sequences of the form
\beqa
t\mapsto C_{1,t}\ldots C_{n,t}\one_{\De}(U), \label{QFT-propagation-observable}
\eeqa
where $\De\subset G_{2m}$ is an open set, whose extension is small compared to $m$, and  
$C_{i,t}$, $i=1,\ldots,n$ are defined as in  (\ref{C-t-observable}). So far we can 
handle detectors which satisfy the following admissibility conditions:
\begin{enumerate}

\item[(a)] The energy-momentum transfers of $B_i^*$ are small neighbourhoods of distinct points 
$p_i$  on the mass hyperboloid  $H_m$ s.t. $ p_1+\cdots+ p_n\in \De$.

\item[(b)] The functions $h_{i,t}\in \cS(T^*\real^d)$ have the form
\beqa
h_{i,t}(x,\xi)=h_{0,i}(x/t)\chi(x/t-\nabla\om(\xi)), \non
\eeqa 
where $h_{0,i}\in C_0^{\infty}(\real^d)$ have mutually disjoint supports, $\chi\in C_0^{\infty}(\real^d)$ is supported
in a small neighbourhood of zero and $\om(\xi):=\sqrt{\xi^2+m^2}$.

\end{enumerate}
The first assumption above says that $B_i^*$ are `creation operators' of single-particle states with energy-momentum
vectors centered around $p_i$. The disjointness of supports of $h_{0,i}$ in the second assumption
ensures that the corresponding detectors are localized in spacelike separated regions for large $t$,
which helps to exploit locality. The function $\chi$ keeps the average velocity $x/t$   close to the
instantaneous velocity $\nabla\om(\xi)$ in accordance with the expected ballistic motion of a particle at asymptotic
times. 
\bet\label{Main-result} Let $\De\subset G_{2m}$ be a small open subset,
$B_i,h_i$, $i=1,\ldots,n$, be admissible as specified in \emph{(a)}, \emph{(b)} above and let $C_{i,t}$ be given by (\ref{C-t-observable}).
Then there exists the limit
\beqa
Q^+_n(\De):=\slim_{t\to\infty}C_{1,t}\ldots C_{n,t}\one_{\De}(U). 
\label{asymptotic-observable}
\eeqa
\eet

\section{Outline of the proof of Theorem~\ref{Main-result}}
\setcounter{equation}{0}

The proof of Theorem~\ref{Main-result} given in \cite{DG13} relies on the method of propagation 
estimates combined with the Haag--Ruelle scattering theory. Here we outline a different argument 
(also obtained jointly with C. G\'erard) which does not use the  Haag--Ruelle theory. 

We note that for $n=1$  condition (a)  is not compatible with $\De\subset G_{2m}$, hence the theorem does not provide any information on the convergence of one detector.  Let us consider the first interesting case which is $n=2$:
Making use of locality and disjointness of supports of $h_{0,1}$, $h_{0,2}$ (condition (b))
we can write  
\beqa
C_{1,t}C_{2,t}\one_{\De}(U)=\e^{\i tH} a_{B_1,B_2}^*(\one_{\hil}\otimes h_{1,t}^\w h_{2,t}^\w)a_{B_1,B_2}\e^{-\i tH}\one_{\De}(U)+O(t^{-\infty}), \label{condition-b-used-first-time}
\eeqa
where $O(t^{-\infty})$ denotes a term which vanishes in norm faster than
any inverse power of $t$. Thus it is enough to show strong convergence of
\beqa
f(t):=\e^{\i tH} a_{B_1,B_2}^*(\one_{\hil}\otimes h_{1,t}^\w h_{2,t}^\w)a_{B_1,B_2}\e^{-\i tH}\one_{\De}(U)
\eeqa 
as $t\to\infty$. The key observation, which allows us to transport methods from quantum-mechanical
scattering theory to the present context, is that the time-derivative of $f$ can be expressed by the Heisenberg derivative of
the quantum-mechanical propagation observable $t\mapsto h_{1,t}^\w h_{2,t}^\w$. More precisely
\beqa
\pa_t f(t)=\e^{\i tH} a_{B_1,B_2}^*(\one_{\hil}\otimes \mathcal{D}(h_{1,t}^\w h_{2,t}^\w))a_{B_1,B_2}\e^{-\i tH}\one_{\De}(U)
+O(t^{-\infty}), \label{time-derivative-property}
\eeqa
where $\mathcal{D}=\pa_t+\i [\ti\om, \,\cdot\,]$ and $\ti\om=\om(-\i\nabla_{x_1})+\om(-\i\nabla_{x_2})$.
Thus the asymptotic time-evolution of a relativistic QFT is governed by a quantum-mechanical Hamiltonian $\ti\om$.
In the case of $n=2$, using relation~(\ref{time-derivative-property})  and the standard phase space propagation estimates (see \emph{e.g.} \cite{DG97}) one can show the convergence of $t\mapsto f(t)$ \cite{DG12}. In this case  functions
 $\chi$ appearing in condition (b) above are not needed and one can prove the convergence of a product of two conventional
Araki--Haag detectors (\ref{Araki-Haag}).  For $n>2$ things become more complicated due to our limited understanding of
quantum mechanical \emph{dispersive systems}, that is systems of particles with non-quadratic dispersion relations. In this
case the phase space propagation estimate is not available and we had to derive a new propagation estimate to control
the convergence of $t\mapsto f(t)$. As explained in more detail in \cite{DG13}, it requires the presence of the functions $\chi$ in the propagation observables.

Let us conclude this section with a few remarks about the  key property (\ref{time-derivative-property}): The term involving 
$\pa_t (h_{1,t}^\w h_{2,t}^\w)$ is self-explanatory. Let us indicate how $\pa_t \e^{\i Ht}$ and $\pa_t \e^{-\i Ht}$ give rise to 
the commutator with $\ti\om$: We note the identities
\beqa
B_1(x_1)B_2(x_2)\one_{\De}(U)&=\one_{\{0\}}(U)B_1(x_1)\one_{H_m}(U) B_2(x_2)\one_{\De}(U), \label{energy-momentum-transfer-one}\\
B_1(x_1)B_2(x_2) H\one_{\De}(U)&=HB_1(x_1)B_2(x_2)\one_{\De}(U)+[B_1(x_1),[B_2(x_2), H]]\one_{\De}(U) \label{simple-algebra}\\
&\phantom{44}+[B_1(x_1),H]B_2(x_2) \one_{\De}(U)+[B_2(x_2), H] B_1(x_1)\one_{\De}(U), \non
\eeqa
where (\ref{energy-momentum-transfer-one}) follows from condition (a) and (\ref{energy-momentum-transfer}), and (\ref{simple-algebra})
is a simple computation. The first term on the r.h.s. of (\ref{simple-algebra}) vanishes due to (\ref{energy-momentum-transfer-one}) and
translation invariance of the vacuum. The term involving the double commutator contributes to $O(t^{-\infty})$ on the r.h.s. of (\ref{time-derivative-property}) due to locality and the disjointness of supports of $h_{1,0}$ and $h_{2,0}$. The last
two terms on the r.h.s. (\ref{simple-algebra}) give rise to the quantum-mechanical Hamiltonian $\ti\om$. In fact, keeping (\ref{energy-momentum-transfer-one}) in mind, we note that
\beqa\bes
\one_{\{0\}}(U)[B_i(x_i),H] \one_{H_m}(U)&=\one_{\{0\}}(U) B_i(x_i)\om(P)\one_{H_m}(U)\\
&=\om(-\i\nabla_{x_i})\one_{\{0\}}(U) B_i(x_i)\one_{H_m}(U),
\ees\eeqa 
where  the last step follows by taking the Fourier transform of $\om$ and exploiting the invariance of the vacuum.

\section{The problem of asymptotic completeness}
\setcounter{equation}{0}

We recall that under the assumptions from Section~\ref{Framework} the  
Haag--Ruelle scattering theory \cite{Ha58, Ru62} 
gives a canonical wave operator  whose range will be denoted by $\hil^+$.  $\hil^+$ can equivalently be seen as a subspace of $\hil$ spanned by (outgoing) scattering states, including   the vacuum ($0$-particle state) and $\hil_1:=\one_{H_m}(U)\hil$ (single-particle states). The property of asymptotic completeness requires that $\hil^+=\hil$ \emph{i.e.}  that all states in the physical Hilbert space can be interpreted as configurations of particles. It is well known that this property does
not follow from the Haag--Kastler postulates: Firstly, it could happen that not all the superselection sectors of the 
theory are accommodated in $\hil$. Single-particle states of charged particles from the omitted sectors  would then  
be missing in $\hil$ and consequently their scattering states would not belong to $\hil^+$. But scattering configurations
of such particles with  total charge zero would clearly belong to $\hil$ thus violating asymptotic completeness. 
Secondly, even if all the   superselection sectors are accommodated in $\hil$,
the physical Hilbert space may still contain pathological states  with too many local degrees  
of freedom which do not admit any particle interpretation. Such states appear \emph{e.g.} in certain generalized free fields \cite{Gre61}.

One approach to the problem of asymptotic completeness in AQFT is to amend the Haag--Kastler postulates with 
some physically motivated \emph{a priori} conditions which should characterize theories with a reasonable particle interpretation.
This approach, pioneered in \cite{HS65}, resulted in a family of \emph{phase space conditions} which brought many important 
insights in AQFT, but did not have much impact on its scattering theory. Our strategy is different: we try to get as close 
as possible to proving complete particle interpretation  without adopting  additional assumptions. \emph{A posteriori} our results can be reformulated as a condition for asymptotic completeness.   
 
To illustrate this strategy, let us discuss the following result from \cite{DG13} which complements Theorem~\ref{Main-result}:
\bet\label{complement} Under the assumptions of Theorem~\ref{Main-result}, the range of each asymptotic observable $Q^+_n(\De)$
belongs to the subspace of scattering states $\one_{\De}(U)\hil^+$.  Moreover, $\one_{\De}(U)\hil^+$ is spanned
by the ranges of $Q^+_n(\De)$, $n\in\nat$.
\eet
We note that the second part of this result ensures that `sufficiently many' asymptotic observables, constructed in Theorem~\ref{Main-result} are non-zero. This part is relatively easy to prove proceeding similarly as in \cite{AH67}. 
The essential part of Theorem~\ref{complement} is the first statement, which says that a certain subspace of $\hil$
(namely the span of the ranges of the asymptotic observables  $Q^+_n(\De)$), which \emph{a priori} has nothing to do with scattering
states, is in fact contained in the subspace of scattering states $\hil^+$. Thus the requirement that the ranges of  $Q^+_n(\De)$ (together with  the vacuum vector and single-particle states) span the entire Hilbert space $\hil$, is a condition for asymptotic completeness.

If we are ready to interpret $Q^+_n(\De)$ as  particle detectors, Theorem~\ref{complement} has a simple physical interpretation. It says that for any initial  state $\Psi\in \hil$, the state resulting from the measurement 
$Q^+_n(\De)$  will be a vector in $\hil^+$, that is a configuration of particles. Theorem~\ref{complement} also says that if $\Psi$ belongs to the orthogonal complement of $\hil^+$ (as a neutral configuration of particles from superselection sectors omitted in $\hil$,
a state with too many local degrees of freedom or any other pathology 
Haag--Kastler postulates may admit) then
$\Psi$ will be  annihilated by  $Q^+_n(\De)$.   This insight cannot be obtained by  methods from \cite{AH67}, where only the existence of  asymptotic observables on $\hil^+$ is considered.

In the  language of many-body scattering theory our results concern the problem of asymptotic completeness
in the \emph{free region} (\emph{i.e.} outside of the collision planes $ \{\,x_1=x_2\}$, $\{x_1=x_3\}$, etc.) In fact,
condition (b) above  ensures that the classical propagation observables, corresponding to (\ref{QFT-propagation-observable}), vanish near the collision planes. 
A natural future direction  is to drop condition (b) and understand the asymptotic dynamics
of AQFT close to the collision planes. As this condition  enters already in the first step of our analysis 
(see (\ref{condition-b-used-first-time})), this goal surely requires  new ideas. Taking quantum mechanics as 
a guide, some relativistic counterpart of the Mourre theory  may be needed here. One can speculate that the covariance under Lorentz boosts, which we did not exploit in our analysis so far, will provide this missing ingredient.

\end{document}